\documentclass[prl,twocolumn,superscriptaddress,preprintnumbers]{revtex4}
\usepackage{graphicx}  
\usepackage{dcolumn}   
\usepackage{bm}        
\usepackage{amssymb}   

\usepackage{comment}
\usepackage{array,multirow,graphicx}

\providecommand{\U}[1]{\protect\rule{.1in}{.1in}}
\usepackage{array,hhline,dcolumn}%
\usepackage[normalem]{ulem}	

\usepackage{wrapfig}
\usepackage{amssymb}
\usepackage{amsmath}
\usepackage{amsfonts}

\usepackage[]{units}

\begin{document}



\title{First Measurement of Neutrino and Antineutrino Coherent Charged Pion Production on Argon}

\author{R.~Acciarri}
\affiliation{Fermi National Accelerator Laboratory, Batavia, IL 60510 USA}
\author{C.~Adams}
\affiliation{Yale University, New Haven, CT 06520 USA}
\author{J.~Asaadi}
\affiliation{Syracuse University, Syracuse, NY 13244 USA}
 \author{B.~Baller}
 \affiliation{Fermi National Accelerator Laboratory, Batavia, IL 60510 USA}
 \author{T.Bolton}
 \affiliation{Kansas State University, Manhattan, KS 66506 USA}
 \author{C.~Bromberg}
 \affiliation{Michigan State University, East Lansing, MI 48824 USA}
 \author{F.~Cavanna}
  \affiliation{Yale University, New Haven, CT 06520 USA}
  \affiliation{Universit\`a dell'Aquila e INFN, 67100 L'Aquila, Italy}
  \author{E.~Church}
  \affiliation{Yale University, New Haven, CT 06520 USA}
 \author{D.~Edmunds}
 \affiliation{Michigan State University, East Lansing, MI 48824 USA}
 \author{A.~Ereditato}
 \affiliation{University of Bern, 3012 Bern, Switzerland}
 \author{S.~Farooq}
 \affiliation{Kansas State University, Manhattan, KS 66506 USA}
 \author{B.~Fleming}
\affiliation{Yale University, New Haven, CT 06520 USA}
 \author{H.~Greenlee}
\affiliation{Fermi National Accelerator Laboratory, Batavia, IL 60510 USA}
 \author{R.~Hatcher}
\affiliation{Fermi National Accelerator Laboratory, Batavia, IL 60510 USA}
 \author{G.~Horton-Smith}
 \affiliation{Kansas State University, Manhattan, KS 66506 USA}
 \author{C.~James}
\affiliation{Fermi National Accelerator Laboratory, Batavia, IL 60510 USA}
 \author{E.~Klein}
\affiliation{Yale University, New Haven, CT 06520 USA}
 \author{K.~Lang}
 \affiliation{The University of Texas at Austin, Austin, TX 78712 USA}
 \author{P.~Laurens}
 \affiliation{Michigan State University, East Lansing, MI 48824 USA}
 \author{R.~Mehdiyev}
 \affiliation{The University of Texas at Austin, Austin, TX 78712 USA}
 \author{B.~Page}
 \affiliation{Michigan State University, East Lansing, MI 48824 USA}
 \author{O.~Palamara}
\affiliation{Yale University, New Haven, CT 06520 USA}
\affiliation{INFN - Laboratori Nazionali del Gran Sasso, 67100 Assergi, Italy}
 \author{K.~Partyka}
\affiliation{Yale University, New Haven, CT 06520 USA}
 \author{G.~Rameika}
\affiliation{Fermi National Accelerator Laboratory, Batavia, IL 60510 USA}
 \author{B.~Rebel}
 \affiliation{Fermi National Accelerator Laboratory, Batavia, IL 60510 USA}
  \author{E.~Santos}
 \affiliation{ Imperial College London, London, SW7 2AZ, UK}
 \author{A.~Schukraft}
 \affiliation{Fermi National Accelerator Laboratory, Batavia, IL 60510 USA}
 \author{M.~Soderberg}
 \affiliation{Syracuse University, Syracuse, NY 13244 USA}
 \affiliation{Fermi National Accelerator Laboratory, Batavia, IL 60510 USA}
 \author{J.~Spitz}
\affiliation{Yale University, New Haven, CT 06520 USA}
 \author{A.M.~Szelc}
\affiliation{Yale University, New Haven, CT 06520 USA}
 \author{M.~Weber}
 \affiliation{University of Bern, 3012 Bern, Switzerland}
\author{T.~Yang}
\affiliation{Fermi National Accelerator Laboratory, Batavia, IL 60510 USA}
\author{G.P.~Zeller} 
\affiliation{Fermi National Accelerator Laboratory, Batavia, IL 60510 USA} 

\collaboration{ArgoNeuT Collaboration}

\date{\today}

\preprint{FERMILAB-PUB-14-258-E}

\begin{abstract}
We report on the first cross section measurements for 
charged current coherent pion production by neutrinos and antineutrinos
on argon. These measurements are performed using the ArgoNeuT detector
exposed to the NuMI beam at Fermilab. The cross sections are measured to be
$
2.6^{+1.2}_{-1.0}(stat)^{+0.3}_{-0.4}(syst) \times 10^{-38} \textrm{cm}^{2}/\textrm{Ar}
$
for neutrinos at a mean energy of $\unit[9.6]{GeV}$ and
$ 
5.5^{+2.6}_{-2.1}(stat)^{+0.6}_{-0.7}(syst) \times 10^{-39} \textrm{cm}^{2}/\textrm{Ar}
$
for antineutrinos at a mean energy of $\unit[3.6]{GeV}$.
\end{abstract}

\pacs{ 25.30.Pt,13.15.+g}
\maketitle

Neutrinos can produce single pion final states by coherently scattering 
from the entire nucleus. Both neutral current (NC) and charged 
current (CC) processes are possible. In these interactions, the squared four-momentum transfer 
to the target nucleus, $\left | t \right |$, is small so the 
nucleus remains unchanged. In this Letter, we focus on the CC 
coherent pion production from muon neutrinos and antineutrinos on argon:
\begin{align}
\nu_{\mu} + \textrm{Ar} &\rightarrow \mu^{-}+\pi^{+}+\textrm{Ar}; \label{eqn1} \\
\bar{\nu}_{\mu} + \textrm{Ar} &\rightarrow \mu^{+}+\pi^{-}+\textrm{Ar}; \label{eqn2}
\end{align}
where the low $\left | t \right |$ condition entails that the pions and 
muons are forward going with respect to the incoming neutrino direction.

There are several models from which one can extract cross sections 
and kinematical predictions for this interaction. 
The Rein-Seghal~\cite{rein} model has been used to successfully 
describe high energy data within experimental uncertainties since the first 
observation of coherent pion production at the Aachen-Padova spark 
chamber~\cite{padova} in 1983. This approach is based on Adler's Partially Conserved Axial Current (PCAC) 
theorem~\cite{adler}, which relates the pion production cross 
section to the cross section for the pion-nucleus scattering. This model 
is still the standard for neutrino generators today, such as {\sc genie}~\cite{genie}, 
{\sc nuwro}~\cite{nuwro}, and {\sc neut}~\cite{neut}, with continued updates to the formalism 
and the pion-nucleus scattering data that is used. With 
recent interest in coherent pion production in the theoretical community, 
other PCAC models 
have been proposed~\cite{berger, paschos}. Microscopic models~\cite{alvarez_ruso,hernandez,nakamura} have also been suggested, which 
employ a full quantum mechanical treatment that explores 
the excitation and decay of the $\Delta$ resonance. While the 
PCAC based models are a simple approach, tailored for the description 
of high energy data, their extension to the few GeV regime is 
not straightforward. Notably, the K2K~\cite{K2K} and SciBooNE~\cite{SciBooNE} 
collaborations found cross section upper limits for the CC coherent 
pion production well below Rein-Seghal's estimation.
The microscopic models are better motivated 
at lower neutrino energies but currently cannot be used to 
describe high energy data. Given the differences in these models, 
more experimental 
measurements are necessary to validate and tune the models and, 
in particular, better understand the transition region between 
microscopic and PCAC validity at $E_{\nu}\sim\unit[3-5]{GeV}$.

In this Letter, a measurement of CC coherent pion production 
from the ArgoNeuT (Argon Neutrino Test) experiment is presented. 
ArgoNeuT~\cite{flavio} is a 
$\unit[170]{L}$  liquid argon 
time projection chamber (LArTPC), with dimensions $47\times40\times\unit[90]{cm^{3}}$. 
The electric field inside the TPC is $\unit[481]{V/cm}$, and the 
drifted charge from particle interactions is read out 
in two planes of 240 wires with $\unit[4]{mm}$ pitch (the induction and collection planes). 
The angle between the induction and collection plane wires is 60 
degrees. 
ArgoNeuT is exposed to the NuMI beam~\cite{numibeam} set 
in an antineutrino-enhanced mode, which provides a flux that is mostly 
muon antineutrino but still rich in muon neutrinos. The total number of protons on target (POT) 
accumulated during a 5-month run is $1.2\times10^{20}$ and the estimated integrated fluxes are
$6.6\times10^{11}$ muon neutrinos per $\unit[]{cm^{2}}$  and
$3.0\times10^{12}$ muon antineutrinos per $\unit[]{cm^{2}}$. 
The differential flux can be found in reference~\cite{cc_inclusive}.
Neutrino interactions comprise almost $60\%$ of all the neutrino/antineutrino-induced events 
in the detector~\cite{cc_inclusive}. During this run, the MINOS near detector~\cite{minossteel} placed 
downstream of ArgoNeuT is also operational. The muons that exit 
ArgoNeuT's TPC volume are matched to MINOS, in which the momentum and 
charge are reconstructed.

Using the \textsc{LArSoft} software~\cite{larsoft}, (anti)neutrino interactions 
are reconstructed, rendering a full characterization of 
the charged particles emerging in the ArgoNeuT detector. The software also provides the framework 
for a Monte Carlo (MC) simulation of the experiment. This is achieved by employing 
{\sc genie}~\cite{genie} as the neutrino event generator and {\sc geant4}~\cite{geant4} 
for the simulation of the propagation of products in the detector. 
The complete ArgoNeuT geometry is simulated along with the signal formation processes and taking into account electronic noise.
The simulated events are fully reconstructed in the same way as for data.
The propagation of particles in the MINOS near detector is simulated with \textsc{geant3} ~\cite{geant3}. 
A standalone version of MINOS simulation and reconstruction is used to characterize the 
matching of tracks passing from ArgoNeuT into MINOS.

The search for CC coherent pion production starts with an 
event selection which is used to find the
two track topology of Eqs. (1) and (2). 
Each of the selection criteria described below is chosen in order to 
maximise the significance, defined as $s/\sqrt{s+b}$, 
where $s$ and $b$ are the numbers of signal and background events 
which pass the selection in the MC simulation. 
The MC used assumes the signal as modeled by Rein-Seghal.
We start by requiring that two tracks are reconstructed in the event, originating 
from the same vertex. One track, identified as the muon, must be reconstructed 
in both ArgoNeuT and MINOS and matched between the two detectors. 
The unmatched track is the pion candidate. 
ArgoNeuT's precise calorimetry 
is used to discriminate pions from protons by defining an acceptance window for the 
mean $dE/dx$ of the unmatched track. 
While the $dE/dx$ of 
a pion will correspond to a Minimum Ionizing Particle ($\unit[1]{MIP}$), a proton track will leave 
an energy deposition several times higher ($>\unit[2]{MIP}$). By applying a selection criteria on the 
$dE/dx$ of the pion-candidate track, the CC quasi-elastic background is almost fully removed.
The calorimetry capabilities of the detector are further exploited by 
investigating the Analog-to-Digital (ADC) readout at the wires at the vertex. Low energy protons 
emerging at the vertex induce high ADC readouts at the first wire hits which are used to exclude the event. 
This selection reduces the background of interactions where multiple low-energy protons are produced, added either by 
nuclear effects or a result of deep inelastic scattering.

The lack of any particles other than the muon and the pion emerging from the vertex is further reinforced by another selection criteria.
For each event, the charge readout inside a $\sim\unit[20]{cm}\times\unit[20]{cm}$ box 
defined in the wire number versus drift time view of the collection plane is counted; the fraction of this charge that is associated with the two outgoing tracks must amount to at least 
$86\%(84\%)$ for antineutrino(neutrino) events. The collection plane is used because its response is better calibrated compared with the induction plane. This verification is crucial since it removes background events with activities around 
the interaction vertex that are not originated from the muon and the pion. 

The event selection defined makes the most of the precise calorimetry and the high imaging 
resolution the ArgoNeuT detector is capable of, which are a characteristic of LArTPCs.
We estimate the selection efficiencies to be $(18.4\pm1.8)\%$ for neutrino and $(21.8\pm0.8)\%$ for antineutrino events.
The inefficiency is dominated by the track reconstruction inefficiency for overlapping 
tracks or complex topologies when the pion interacts with the argon nucleus. The systematic uncertainties 
associated to the assumptions on the kinematics of the signal events are accessed by estimating the efficiency 
using a different generator (\textsc{nuwro}). The difference between the efficiencies obtained with the two generators is 
kept as the systematic uncertainty.  

\begin{figure}
\includegraphics[width=\columnwidth]{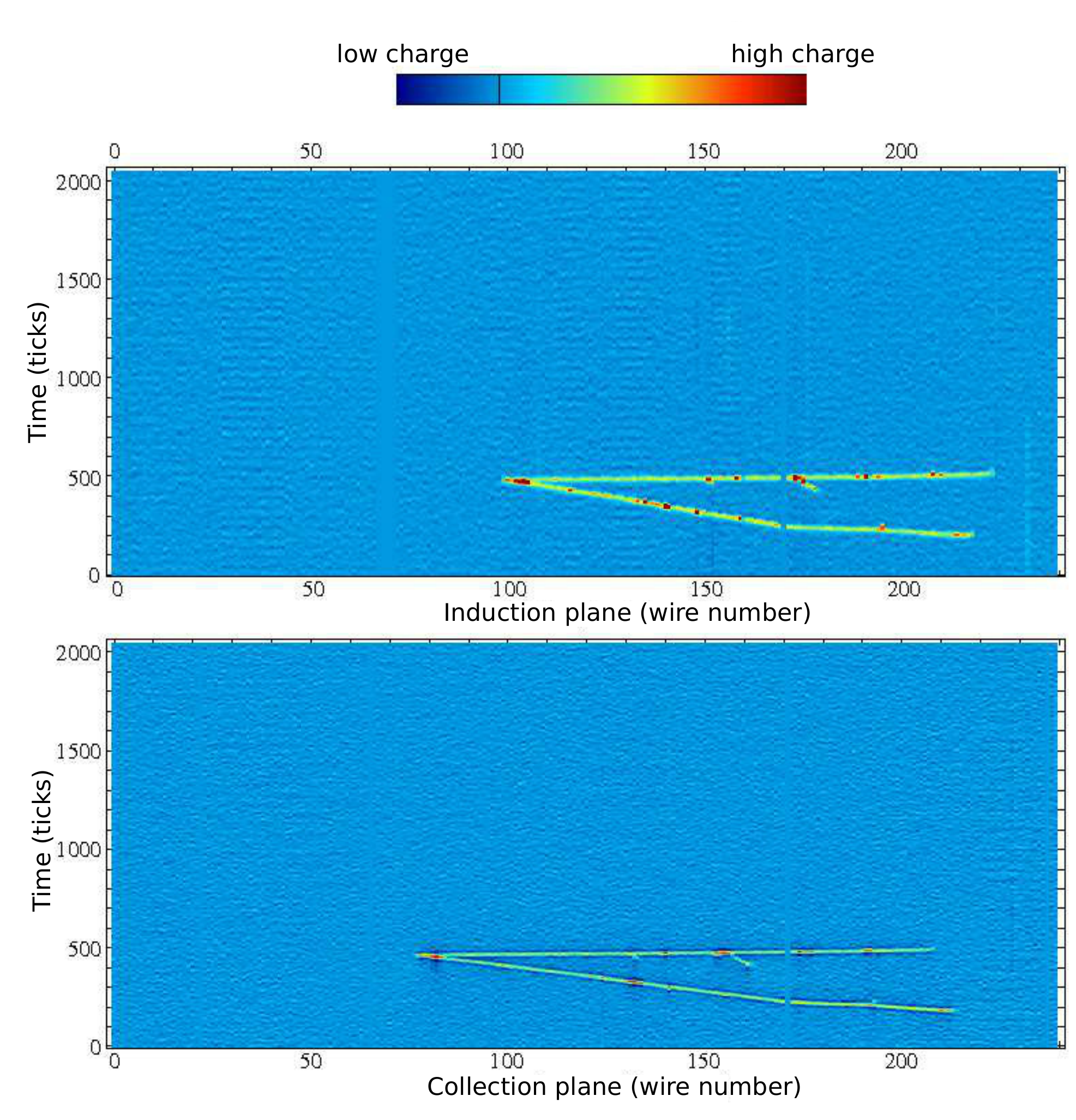}
\caption{\label{fig:evd} An example of CC Coherent 
pion production from a neutrino in ArgoNeuT. The neutrino's incoming direction is 
along the horizontal coordinate; the muon track corresponds to the most forward going one, 
making an angle of $1.2^{\circ}$ with the incoming neutrino direction. The opening angle between 
the muon and the pion track is $10.6^{\circ}$. A kink in the pion trajectory is visible.}
\end{figure}

A total of $167$ antineutrino and $150$ neutrino events have the two-track topology in the TPC with one track matched to a reconstructed track in MINOS.  
After applying the event selection described, $30$ antineutrino and $24$ neutrino candidate events remain.
This event sample contains a background fraction, predominantly resonant and deep inelastic interactions, 
that ideally would be reduced by selecting events with low $\left | t \right |=\left | (q-p_{\pi})^{2} \right |$, where $q$ represents the 
momentum transfer from the neutrino and $p_{\pi}$ is the momentum carried by the pion. 
This approach is not feasible because most pions are not contained 
in the ArgoNeuT TPC so their momentum can't be estimated. Instead, we achieve signal from background separation 
by applying a multivariate method which exploits 
the topological and calorimetric information reconstructed in each event. The ROOT 
Toolkit for Multivariate Analysis~\cite{tmva} was used to create a 
Boosted Decision Tree (BDT) which is trained using \textsc{genie} signal and background samples. The classification is based on 
the angles of the pion and muon tracks, 
the visible energy loss of the pion from the TPC's calorimetry, 
the reconstructed muon momentum from MINOS and
the mean stopping power of the first third of the muon track. 
The last of these parameters was added to help distinguish events where the start of 
the muon and pion tracks is overlapping. The angular parameters have the highest discrimination 
power. An example of a neutrino interaction classified as signal by the BDT is show in Figure~\ref{fig:evd}.

\begin{figure}
\includegraphics[width=\columnwidth]{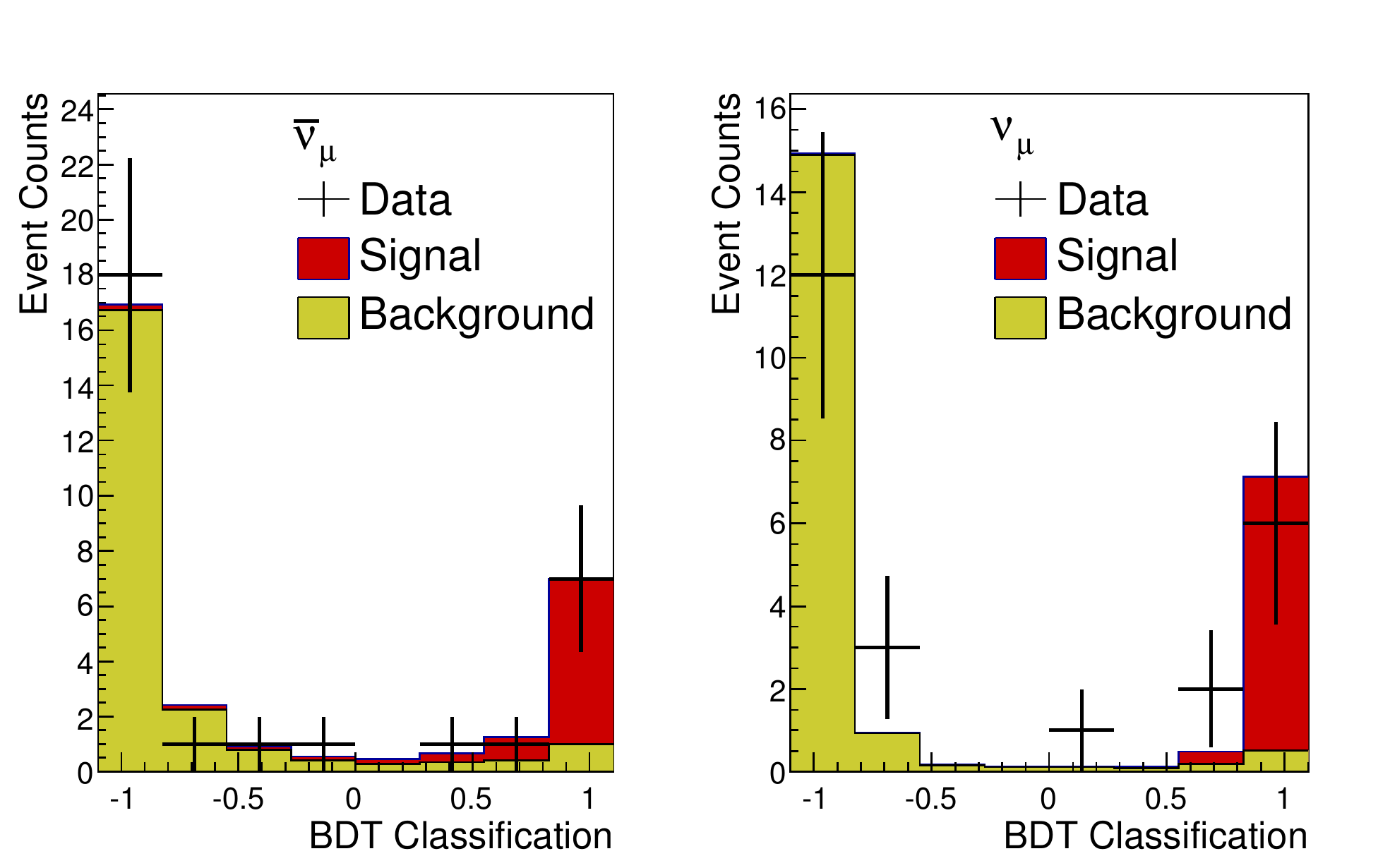}
\caption{\label{fig:fit} Best-fit of the signal and background templates to the data. 
A BDT classification value of $-1$ means background-like and $1$ is signal-like. 
The background and signal shapes are 
scaled to minimize an effective $\chi^{2}$ function from which the statistical error is also extracted.}
\end{figure}

To estimate the rate of signal events, the BDT distribution in data is fitted to a linear 
combination of templates for signal and background obtained from simulation.
The fit preserves the shape of the signal and background BDT distributions and finds the scale of these 
which best agrees with the data by minimising the effective $\chi^{2}=-2\ln \mathcal{L}$, where $\mathcal{L}$ represents 
the likelihood assuming Poisson-distributed counts in each bin.
The statistical error is found by evaluating the $1\sigma$ 
interval, determined by $\Delta \chi^{2}=\chi^{2}-\chi^{2}_{min}=1$. 
Figure~\ref{fig:fit} shows the data and the best-fit signal and background distributions. 
The antineutrino signal is estimated to be $7.9_{-3.0}^{+3.7}$ events and the neutrino signal is $7.0_{-2.6}^{+3.3}$ events. 
The background contamination in the signal region (BDT Classification $>$ 0) is small. 

The systematic uncertainties affecting the measurement are listed in Table~\ref{tab:syst}.
These are dominated by the flux-scale uncertainty ($10-12\%$). Reconstruction effects have their impact
estimated by adjusting the reconstructed values by $\pm 1 \sigma$, where $\sigma$ is the 
uncertainty on the reconstructed parameter. 
The absolute muon momentum estimated from the track curvature in the MINOS detector has a 4\% systematic uncertainty~\cite{minos_cc_inclusive} 
and the angular uncertainty assigned to tracks reconstructed in ArgoNeuT is $1^{\circ}$~\cite{spitz}.
The contribution of background uncertainties is found by adjusting the contribution 
from each individual background process by $\pm20\%$ ~\cite{jorge}.
The rate at which the charge of the muon is mis-identified is also 
estimated and treated like the other backgrounds, though its contribution was found to be negligible. 
The effect of nuclear interactions affecting the production of background events is also considered. 
This is done by evaluating the fraction of background events added by final 
state interactions and re-weighting this sample by a conservative factor ($\pm20\%$).
Finally, the systematic error associated with the signal modeling is investigated by generating 
a signal template using \textsc{nuwro}. The difference in the number of signal events found after 
repeating the fit is our estimation of the systematic uncertainty.

\begin{table}
\caption{\label{tab:syst}Contributions to the total systematic uncertainty on the flux-averaged 
cross sections. The dominant backgrounds in this analysis are the 
CC quasi-elastic (QE), resonant (RES), and deep inelastic scattering (DIS) interactions.}

\begin{ruledtabular}
\begin{tabular}{lcc}
\multicolumn{3}{r}{Cross Section Uncertainty [\%]} \\ 
\hline
Systematic Effect & $\bar{\nu}_{\mu}$ & $\nu_{\mu}$ \\
\hline
Flux normalization  & $^{+10.0}_{-12.0}$ & $^{+10.0}_{-12.0}$ \\
\hline
MINOS momentum res. & $\pm4.1$ & $\pm4.3$     \\
\hline
ArgoNeuT angle res. & $\pm1.6$           & $\pm2.7$ \\
\hline
CC QE background      & $^{+0.3}_{-0.4}$ & $^{+1.2}_{-0.6}$ \\
\hline 
CC RES background    & $^{+0.2}_{-0.5}$ & $^{+0.4}_{-0.3}$  \\
\hline
CC DIS background    & $\pm0.1$  & $\pm0.3$  \\
\hline
Nuclear Effects & $\pm0.3$ & $\pm0.7$ \\
\hline
POT           & $\pm0.1$ & $\pm0.1$ \\
\hline
Number of Argon Targets & $\pm2.2$ & $\pm2.2$ \\
\hline
Efficiency & $\pm0.8$ & $\pm1.8$ \\
\hline
Signal modeling     & $\pm0.8$ & $\pm 5.7$ \\
\hline
Total systematics    & $^{+11.3}_{-13.1}$ & $^{+12.9}_{-14.5}$ \\
\end{tabular}
\end{ruledtabular}
\end{table}

\begin{figure}[t]
  \includegraphics[width=\columnwidth]{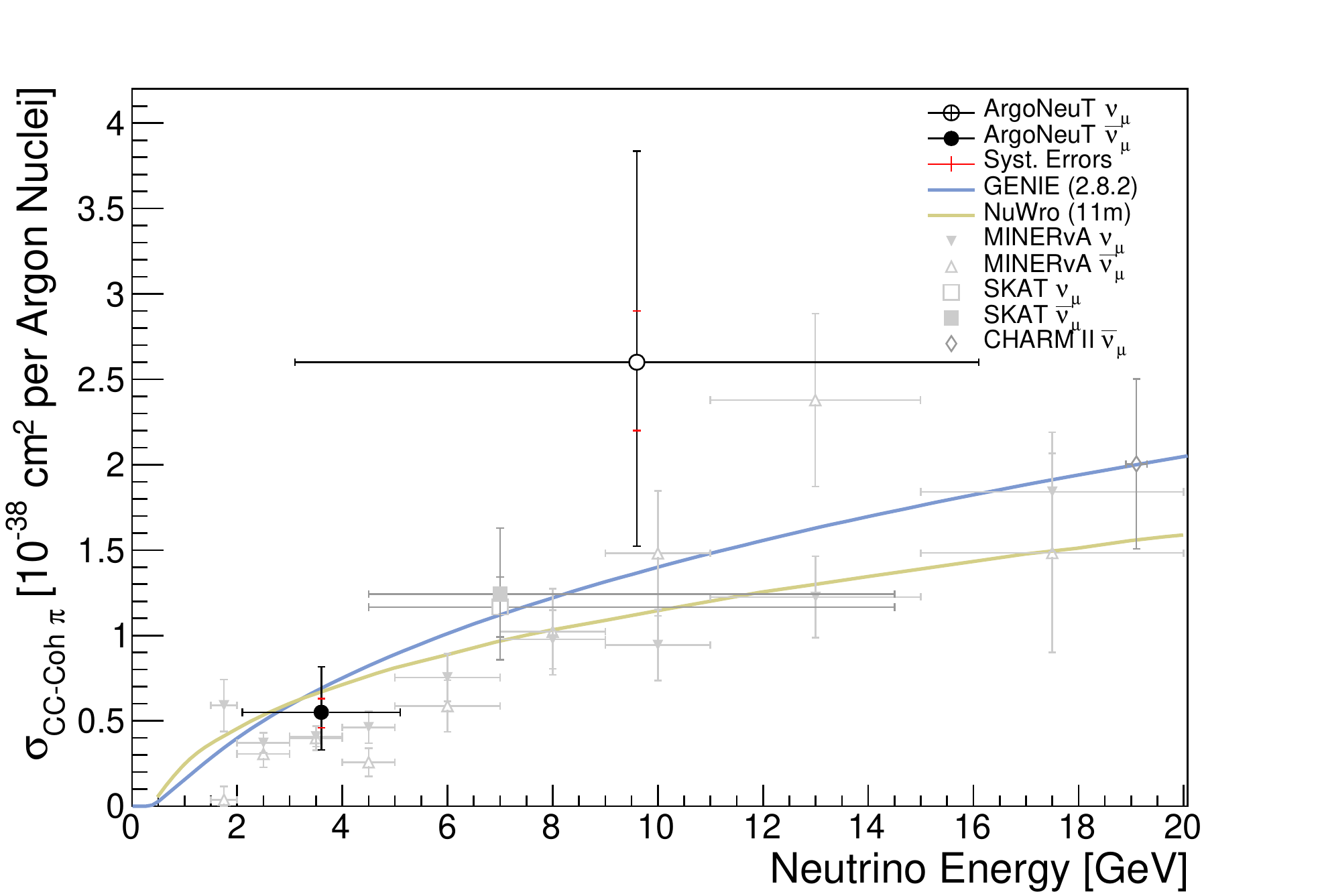}
  \caption{ArgoNeuT's CC coherent pion cross section measurements (open circle and filled circle) 
  compared to Rein-Seghal's model as implemented in {\sc genie} and {\sc nuwro}~\cite{nuwro}.
  The statistical error is dominant (the systematic 
  uncertainty is shown alone for comparison).
  Data from other experiments in the 
  same energy range is also shown. These consist in measurements made by SKAT 
  (filled square, open square), CHARM II (diamond) and MINERvA (triangle, inverted triangle)~\cite{skat_measurement, charm_measurement, minerva_measurement}. 
  These measurements are scaled to Argon assuming the $A^{1/3}$ dependance 
  from the Rein-Seghal model.}
 \label{genie}
\end{figure}

The flux-averaged cross section is found 
by dividing the number of signal events 
by the efficiency of the selection, 
the number of target nuclei in the fiducial volume 
and the integrated (anti)neutrino flux. 
The measurements we report are
\begin{align}
\left < \sigma_{\bar{\nu}_{\mu}} \right > &= 5.5^{+2.6}_{-2.1}(stat)^{+0.6}_{-0.7}(syst) \times 10^{-39} \textrm{cm}^{2} \\
\left < \sigma_{\nu_{\mu}} \right > &= 2.6^{+1.2}_{-1.0}(stat)^{+0.3}_{-0.4}(syst)) \times 10^{-38} \textrm{cm}^{2}
\end{align}
per argon nuclei at $\left<E_{\bar{\nu}_{\mu}}\right> = 3.6\pm1.5$ GeV and $\left<E_{\nu_\mu}\right> = 9.6\pm6.5$ GeV, where the $\pm1.5(6.5)$ GeV represents the range that contains 68\% of the flux. A comparison between these measurements, existing data,
and the Rein-Seghal model are shown in Figure \ref{genie}. The antineutrino measurement
agrees well with the Rein-Seghal model while the neutrino one deviates by $\sim1.2\sigma$.

In this Letter, we have presented the first cross section measurement of CC coherent 
pion production on argon. This is also the first time that machine 
learning techniques have been applied to LArTPC data analysis.
The large uncertainties on the final cross section values are dominated by the 
statistical errors. Using the precise calorimetry and the high resolution of the interaction vertex  
which are fundamental for this analysis, future LArTPC experiments will be able to 
provide decisive measurements for the understanding of neutrino induced coherent 
pion production.

We gratefully acknowledge the cooperation of the MINOS collaboration in providing their data for 
use in this analysis. We wish to acknowledge the support of Fermilab, the Department of Energy, 
and the National Science Foundation in ArgoNeuT's construction, operation, and data analysis.
We also wish to acknowledge the support of STFC, FNAL and FCT's grant SFRH/BD/69814/2010.


\begin{thebibliography}{99}

\bibitem{rein}
\bibinfo{author}{D.~Rein, and L.~M.~Seghal},
  \bibinfo{journal}{Phys. Lett. B} \textbf{\bibinfo{volume}{657}},
    \bibinfo{pages}{207--209} (\bibinfo{year}{2007}).
    
\bibitem{padova}
\bibinfo{author}{H. Faissner, \textit{et al.}},
  \bibinfo{journal}{Phys. Lett. B} \textbf{\bibinfo{volume}{125}},
    \bibinfo{pages}{230--236} (\bibinfo{year}{1983}).

\bibitem{adler}
\bibinfo{author}{S.~Adler},
  \bibinfo{journal}{Phys. Rev.} \textbf{\bibinfo{volume}{135}},
    \bibinfo{pages}{B963--B966} (\bibinfo{year}{1964}).
 
\bibitem{genie}
\bibinfo{author}{C.~Andreopoulos, \textit{et al.}},
  \bibinfo{journal}{Nucl. Instr. \& Meth. A} \textbf{\bibinfo{volume}{614}},
  \bibinfo{pages}{87} (\bibinfo{year}{2010}).

\bibitem{nuwro}
\bibinfo{author}{T.~Golan, C.~Juszczak and J.~T.~Sobczyk},
  \bibinfo{journal}{Phys. Rev. C} \textbf{\bibinfo{volume}{86}},
    \bibinfo{pages}{015505} (\bibinfo{year}{2012}).

\bibitem{neut}
\bibinfo{author}{Y.~Hayato},
  \bibinfo{journal}{Acta Phys. Polon. B} \textbf{\bibinfo{volume}{40}},
    \bibinfo{pages}{2477-2489} (\bibinfo{year}{2009}).
 
\bibitem{berger}
\bibinfo{author}{C.~Berger, L.~M.~Seghal},
  \bibinfo{journal}{Phys. Rev. D} \textbf{\bibinfo{volume}{79}},
  \bibinfo{pages}{053003}
  (\bibinfo{year}{2009}).

\bibitem{paschos}
\bibinfo{author}{E.~A.~Paschos and D.~Schalla},
  \bibinfo{journal}{Phys. Rev. D} \textbf{\bibinfo{volume}{80}},
  \bibinfo{pages}{033005},
   (\bibinfo{year}{2009}).

\bibitem{alvarez_ruso}
\bibinfo{author}{L.~Alvarez-Ruso, L.~S.~Geng, S.~Hirenzaki and M.~J.~VicenteVacas},
  \bibinfo{journal}{Phys. Rev. C} \textbf{\bibinfo{volume}{75}},
  \bibinfo{pages}{055501}
   (\bibinfo{year}{2007}).
    
\bibitem{hernandez}
\bibinfo{author}{E.~Hernandez, J.~Nieves and M.~Valverde},
  \bibinfo{journal}{Phys. Rev. D} \textbf{\bibinfo{volume}{76}},
  \bibinfo{pages}{033005}
   (\bibinfo{year}{2007}).

\bibitem{nakamura}
\bibinfo{author}{S.~X.~Nakamura, \textit{et al.}},
\eprint{arXiv:0901.2366}.
    
\bibitem{K2K}
\bibinfo{author}{M.~Hasegawa, \textit{et al.}},
  \bibinfo{journal}{Phys. Rev. Lett.} \textbf{\bibinfo{volume}{95}},
    \bibinfo{pages}{252301} (\bibinfo{year}{2005}).  

\bibitem{SciBooNE}
\bibinfo{author}{K.~Hiraide, \textit{et al.}},
  \bibinfo{journal}{Phys. Rev. D} \textbf{\bibinfo{volume}{78}},
    \bibinfo{pages}{112004} (\bibinfo{year}{2008}).

\bibitem{flavio}
\bibinfo{author}{C.~Anderson, \textit{et al.} (ArgoNeuT Collaboration)}
  \bibinfo{journal}{JINST} \textbf{\bibinfo{volume}{7}},
  \bibinfo{pages}{P10019} (\bibinfo{year}{2012}).

\bibitem{numibeam}
\bibinfo{author}{K.~Anderson, \textit{et al.}},
  \bibinfo{journal}{FERMILAB-DESIGN-1998-01}  (\bibinfo{year}{1998}).

\bibitem{cc_inclusive}
\bibinfo{author}{R.~Acciarri, \textit{et al.} (ArgoNeuT Collaboration)},
  \bibinfo{journal}{Phys. Rev. D}  
    \textbf{\bibinfo{volume}{89}}, \bibinfo{pages}{112003} (\bibinfo{year}{2014}).

\bibitem{minossteel}
\bibinfo{author}{D.~G.~Michael, \textit{et al.} (MINOS
  Collaboration)}, \bibinfo{journal}{Nucl. Instr. \& Meth. A}
  \textbf{\bibinfo{volume}{596}}, \bibinfo{pages}{190} (\bibinfo{year}{2008}).
  
\bibitem{larsoft}
\bibinfo{author}{E.~Church}, 
\eprint{arXiv:1311.6774}.
  
\bibitem{geant4}
\bibinfo{author}{S.~Agostinelli, \textit{et al.}},
  \bibinfo{journal}{Nucl. Instr. \& Meth. A} \textbf{\bibinfo{volume}{506}},
  \bibinfo{pages}{250} (\bibinfo{year}{2003}).

\bibitem{geant3}
\bibinfo{author}{Application Software Group},
  \bibinfo{journal}{CERN Program Library Long Writeup W5013, CERN},
   \bibinfo{year}{1994}.

\bibitem{tmva} 
\bibinfo{author}{A.~Hocker, \textit{et al.}},
  \bibinfo{journal}{PoS} \textbf{\bibinfo{volume}{ACAT}},
  \bibinfo{pages}{040} (\bibinfo{year}{2007}).
 
\bibitem{minos_cc_inclusive}
\bibinfo{author}{P.~Adamson, \textit{et al.}},
  \bibinfo{journal}{Phys. Rev. D} \textbf{\bibinfo{volume}{81}},
  \bibinfo{pages}{072002} (\bibinfo{year}{2010}).

\bibitem{spitz}
\bibinfo{author}{J.~Spitz},
  \bibinfo{journal}{Ph.D. thesis, Yale University, [FERMILAB-THESIS-2011-36]} (\bibinfo{year}{2011}).

\bibitem{jorge}
\bibinfo{author}{J.~G.~Morfin, and J.~Nieves, and J.~T.~Sobczyk},
  \bibinfo{journal}{Adv. High Energy Phys.} \textbf{\bibinfo{volume}{2012}},
  \bibinfo{pages}{934597} (\bibinfo{year}{2012}).

\bibitem{skat_measurement}
\bibinfo{author}{H.-J.~Grabosch, \textit{et al.}},
  \bibinfo{journal}{Zeitschrift fur Physik C Particles and Fields} \textbf{\bibinfo{volume}{31}},
   \bibinfo{pages}{203-211} (\bibinfo{year}{1986}).


\bibitem{charm_measurement}
\bibinfo{author}{P.~Vilain \textit{et al.}},
  \bibinfo{journal}{Phys. Lett. B} \textbf{\bibinfo{volume}{313}},
   \bibinfo{pages}{267-275} (\bibinfo{year}{1993}).

\bibitem{minerva_measurement}
\bibinfo{author}{A.~Higuera, \textit{et al.}},
  \bibinfo{journal}{Phys. Rev. Lett.} \textbf{\bibinfo{volume}{113}},
    \bibinfo{pages}{261802} (\bibinfo{year}{2014}).  

\end{thebibliography}

\end{document}